\documentclass[english]{article}
\usepackage[T1]{fontenc}
\usepackage[latin9]{inputenc}
\usepackage{amsthm}
\usepackage{amsmath}
\usepackage{amssymb}
\usepackage{wasysym}

\makeatletter
\numberwithin{figure}{section}
\usepackage{enumitem}		% customizable list environments
  \theoremstyle{definition}
  \newtheorem{problem}{\protect\problemname}[section]
  \theoremstyle{plain}
  \newtheorem{ax}{\protect\axiomname}
  \theoremstyle{definition}
  \newtheorem{defn}{\protect\definitionname}[section]
  \theoremstyle{plain}
  \newtheorem{lem}{\protect\lemmaname}[section]
  \theoremstyle{plain}
  \newtheorem{cor}{\protect\corollaryname}[section]
  \theoremstyle{remark}
  \newtheorem{rem}{\protect\remarkname}[section]
  \theoremstyle{definition}
  \newtheorem{example}{\protect\examplename}[section]
  \theoremstyle{plain}
  \newtheorem{thm}{\protect\theoremname}[section]
  \theoremstyle{remark}
  \newtheorem{claim}{\protect\claimname}[section]
  \theoremstyle{plain}
  \newtheorem{prop}{\protect\propositionname}[section]

\makeatother

\usepackage{babel}
  \providecommand{\axiomname}{Axiom}
  \providecommand{\claimname}{Claim}
  \providecommand{\definitionname}{Definition}
  \providecommand{\examplename}{Example}
  \providecommand{\lemmaname}{Lemma}
  \providecommand{\problemname}{Problem}
  \providecommand{\propositionname}{Proposition}
  \providecommand{\remarkname}{Remark}
\providecommand{\corollaryname}{Corollary}
\providecommand{\theoremname}{Theorem}

\begin{document}

\title{The McDougal Cave and Counting issues}

\author{Sayandeep Khan, Synergy Moon LLC}
\maketitle
\begin{abstract}
In this paper I investigate the problem of tagging elements of a set,
and the elements of those elements, uniquely, when they admit an order,
and two boundary elements are tagged. A heuristic sorting algorithm
is also investigated.
\end{abstract}

\section{Introduction }

\subsection{The initial problem}

I will begin by investigating an imaginary problem, and then show
the use of the solution in practical issues. The problem goes as follows.
\begin{problem}
Tom Sawyer is in the McDougal Cave. The cave is a system of interconnected
tunnels, and rooms. Each room is connected to each other via tunnels,
and inside every room, is a self-contained tunnel system. Each tunnel,
may contain smaller rooms, and mazes. Tom wants to systematically
inspect the cave system to find his treasure, however, he need not
inspect every chamber, every tunnel and every room, and similarly,
he can choose in which order he would explore. Nonetheless, he can
be in only one position at a time. In order to mark which room he
is in, with respect to which tunnel, and in which larger room, in
turn, the tunnel itself is, Tom needs to assign a number to his location.
How could such a numbering scheme work, is the question I would attempt
to answer.
\end{problem}
We will see, that this problem is similar to some issues encountered
in modeling and simulation, especially, those which use blocks (assigned
with certain tasks, selected from a pallet) connected to each other.
I will also present yet another solution of the classic sorting problem.

\subsection{Conway's Numbers - primary decisions}

Conway presented a number system {[}1{]} which can be used for arbitrary
counting. This system is the obvious choice for the counting problem
is because it permits assignment of new tags between any two, and
the construction itself admits an ordered structure which can be extended
to infinity. 

However, this number system does not permit too much freedom of tagging,
then only one number can be created between two, and nor is this number
unique to the generation, nor is the generation independent to order
of generation. Moreover, it is considered that such numbers already
exists. Therefore, we consider a different counting system.

\section{Beyond Conway's Numbers}

In this section, I propose an extension of the usual Conway's Numbers,
and investigate the algebraic structure of them. Numbers generated
this way will be called, in this context, ``Extended Conway's Number''-s,
or ``ECN''-s

\subsection{Axioms}

Before further discussion, it is necessary to declare a few axioms.
\begin{ax}
The Axiom of Sets: Sets exist, and any set, including the null set
has a few attributes, such as size etc. Sets are collection of mathematical
objects, and are objects themselves.
\end{ax}
Sets can be assigned with natural attributes, such as size, disjoint
element count, as well as synthetic attributes, e.g. position in a
sequence, morphism, etc. These attributes are used for ordering sets. 
\begin{ax}
The Axiom of ordering: A collection of Sets can be ordered. That is,
comparative relations between sets and their properties exist. The
number of possible ordering is not restricted. It is possible, two
or more orderings can coincide (i.e. at this point, the requirement
for unique ordering is omitted). If two or more orderings coincide,
then they are equivalent, and existence of one implies the existence
of the other. 
\end{ax}
Two orderings, $a<b<...<c$ and $d<e<...<f$ are equivalent $iff$
$a\equiv d,b\equiv e,...,c\equiv f$ etc. At this moment, the notion
of index is not yet defined. Here the symbol $\equiv$ implies ``the
same as'' (dasselbe, exactly the same, of course the exactness of
similarity depends on the context).

There exists the following order relations:
\begin{enumerate}
\item $\equiv$: exactly the same
\item $=$: equivalent. less strict similarity than previous
\item $<$: not (necessarily) similar
\item $>$: not (necessarily) similar, antisymmetric to the previous
\item $\leq$: not $>$
\item $\geq$: not $<$
\end{enumerate}
The third and fourth relations are transformable to each other, so
are the fifth and sixth.
\begin{ax}
Axiom of Numbers: Numbers are symbolic representation of properties
of sets, and they exist, and can be constructed from sets. Each set
has at least one number associated with it. 
\end{ax}
If the same number can be constructed from two or more sets, then
they are equivalent in that context, and vice versa. A collection
of Number can form a set. Numbers can be constructed from other numbers. 
\begin{ax}
Axiom of Construction: A number can be constructed from a set in different
ways using well behaved functions. A number $N$, constructed from
set $S$, using construction method $\mathfrak{C}$, can be written
as $N=\mathfrak{C}(S)$.
\end{ax}
This axiom forms the basis of the rest of the analysis, where we attempt
to construct numbers, from properties of sets, which assign an order
to them.
\begin{ax}
Axiom of Repetition: Any operation can be applied on any object, an
unlimited number of times, but the outcome may be undefined.
\end{ax}
This axiom requires immediately the following axiom:
\begin{ax}
Axiom of Operation: Set operations, including but not limited to Union,
Subtraction, Intersection etc. exists.
\end{ax}
The behavior of these operations are defined in the Euclidean axioms.
\begin{ax}
The Euclidean Axioms apply.
\end{ax}
The Euclidean axioms deal with the ``largeness'' of an object. Largeness
can have different meanings on different contexts. However, axiom
of sets assume that ``size'', in its intuitive sense, is a natural
property of a set.

\subsection{Ordering and Counting}

Ordering is a natural process, therefore, the analysis can start from
ordered sequence of sets.

\subsubsection{Ordered Sequence of Sets and Comparison to Set}

From Axiom of Ordering, sets assume ordering. Such an ordering constructs
a sequence. The elements of a sequence are disjoint. The construction
of a sequence does not depend on the operations described on the axiom
of operations. 

A sequence does not admit a size, it admits a count. A count will
be defined in due course. An uniquely defined sequence is always ordered,
and assumes an ``ordering rule''. An ordering rule can be an enumeration,
or, alternatively an (well behaved) function. 
\begin{defn}
A collection of objects, brought together without those operations,
which is not a sequence, is a $disarray$, or an $unordnung$. An
unordnung, can be converted to a sequence, with a function (which
is an ordering rule). Such functions are called $ordering\,\, functions$,
or $ordnungsregel$-s
\end{defn}
Two ordered sequences are equivalent, $iff$ they are constructed
from two disarrays, that contain pairwise equivalent elements, and
same ordnungsregel apply on them.

A set, however is constructed using set operations. 
\begin{defn}
If a set is to be constructed from elements which admit the $=$ relation
under certain contexts, yet are distinct elements in a sequence, then
they are handled as distinct elements under set operations. This condition
is called $otherness$ condition or $alteritaet$ condition.
\end{defn}

\subsubsection{Order Label}

Since each of the ordered sets has a number associated to it, from
Axiom of Numbers, those numbers are also therefore ordered. The same
ordering, therefore can be achieved by ordering the numbers associated
with the sets too. Therefore:
\begin{lem}
Numbers are Order Labels of sets. They generate an unique ordering.
\end{lem}
That is, numbers label sets, and those labels can determine the ordering. 
\begin{proof}
The term ``Order Label'', in this context, means a symbol, which
can be ordered to achieve the same ordering as the ordering of sets
themselves. To prove, that numbers are Order Labels, it is sufficient
to show that, $\mathfrak{O}\{N=\mathfrak{C}(S)\}\equiv\mathfrak{O}\{S\}$.
Here, $\mathfrak{O}$ is an ordering, and $\mathfrak{\equiv}$is the
equivalence relation.

Take a collection of sets $S_{a},....,S_{o}$. Let two orderings be
$S_{i}<S_{j}<...<S_{l}$, and $S_{e}<S_{f}<...<S_{g}$. The two are
equivalent iff $S_{i}\equiv S_{e}$, $S_{j}\equiv S_{f}$, $S_{k}\equiv S_{g}$. 

From axiom of numbers, each set can generate an unique number, and
if two sets produce the same number, then they are equivalent, and
vice-versa. That is, $\mathfrak{C}(S_{i})=\mathfrak{C}(S_{e})$, etc
implies $S_{i}\equiv S_{e}$, $S_{j}\equiv S_{f}$, $S_{k}\equiv S_{g}$.
Therefore the two orderings are equivalent, if $\mathfrak{C}(S_{i})=\mathfrak{C}(S_{e})$,
etc holds. 

Under that condition, $\mathfrak{C}(S_{i})$ can be generated from
both $S_{i}$, and $S_{e}$. Consider the order sequence $\mathfrak{C}(S_{i})<\mathfrak{C}(S_{j})<...<\mathfrak{C}(S_{l})$.
This sequence implies therefore both the sequences $S_{i}<S_{j}<...<S_{l}$,
and $S_{e}<S_{f}<...<S_{g}$. 

Therefore, if the order sequence $\mathfrak{C}(S_{i})<\mathfrak{C}(S_{j})<...<\mathfrak{C}(S_{l})$
implies two sequences, then both are equivalent. In other words, all
order sequences generated by a number sequence are equivalent. QED.
\end{proof}
The order labels can be used to assign an index to the elements of
a sequence.

\subsubsection{Count Label }

A sequence admits another measure, namely
\begin{defn}
The Order Label of the last element of an ordered sequence is called
the count of the sequence. 

The count is useful to determine the size of set that is constructed
from the sequence, but this will be explicitly defined later. 

A count label is used for extracting subsequences of a sequence, then
the order label of the last element of the target subsequence is the
count of it. In certain cases, enumerating the order labels of the
each individual element is to be avoided, and is done using count
labels.

A count label is used, among other things, to assign a measure to
the iterations of a repetitive operation. 
\end{defn}

\subsubsection{Code Sequences and Groups}

Numbers are symbols. 
\begin{defn}
A string of symbols is called a codeword. A codeword can be constructed
from set $\mathfrak{S}$ of symbols, taking the elements of the set,
possibly repeated, and composing them together. The complete set of
such words is called language.
\end{defn}
Usually, the symbol $\Sigma$ is used instead of $\mathfrak{S}$,
but here to avoid confusion with the summation sign, we use the symbol
$\mathfrak{S}$.
\begin{defn}
The construction of a code sequence is given by a formal $grammar$,
a set of rules, which transfers one word to another.
\end{defn}
If the words are given, and a grammar exists that transfers two words
to a third, then we arrive at the notion of Groups, and similar structures.
Such structures will tell as important details of the set of numbers
under investigation.

\subsection{Consistence to Euclidean Axioms}

The above discussion defines numbers as (symbols to) labels. However,
with labels, operations can not be defined. It is therefore, necessary
to unify the labels with size. Immediately it is clear that size offers
a natural way of ordering sets, with the previous element of the ordered
sequence being of smaller size than the next one.

\subsubsection{Zahlenaufbau}

Before the discussion may proceed further, certain more analysis is
to be considered. This analysis will be called Zahlenaufbau in this
context. 

Consider the fifth axiom of Euclid: any object is larger than any
part of itself, or, any part of an object is smaller than the whole.
An important corollary is:
\begin{cor}
Since the null set is a subset (``part'') of any set, it is the
smallest set. 
\end{cor}
Axiomatically, it is proposed that any part of any object is included
in the whole. So, if an object $\mathfrak{o_{a}}$ contains another
object $\mathfrak{o_{b}}$, then we write, $\mathfrak{o_{b}}\subset\mathfrak{o_{a}},\mathfrak{o_{b}}<\mathfrak{o_{a}}$.
If $\mathfrak{o_{1}}\subset\mathfrak{o_{2}},\mathfrak{o_{1}}\supset\mathfrak{o_{2}}$,
then $\mathfrak{o_{1}}=\mathfrak{o_{2}}$. 

Now, from the notion of union, one can write, $\mathfrak{o_{b}}\cup\mathfrak{o_{c}}=\mathfrak{o_{a}}$.
Consider the general sequence :

$\mathfrak{o}\cup\mathfrak{k}=\mathfrak{o_{k}},\mathfrak{o_{k}}\cup\mathfrak{k^{\prime}}=\mathfrak{o_{k^{\prime}}},\mathfrak{o_{k^{\prime}}}\cup\mathfrak{k^{\prime\prime}}=\mathfrak{o_{k^{\prime\prime}}}$...
ad petitum. Obviously, $\mathfrak{o_{q}\underset{t\mbox{ times}}{\underbrace{\cup\mathfrak{r}}}}=\mathfrak{o_{s}}\implies\mathfrak{o_{q}}<\mathfrak{o_{s}}$. 

Now, let $\mathfrak{k}=\mathfrak{k^{\prime}}=\mathfrak{k^{\prime\prime}}=\mathfrak{K}\neq\phi$,
$\mathfrak{o}=\phi$, the Null set. Obviously, the largeness of the
Null set is an absolute minimum. The largeness, or ``size'' is represented
by the number Zero. $\mathfrak{K}$ is called $einheit$ from German
for unitness, the property of being one. The number immediately constructed
after the nullset is One, w.r.t. $\mathfrak{K}$, the next number
is Two, w.r.t. $\mathfrak{K}$ , et cetra. The size of each $\underline{\bigcup}\mathfrak{k}$
is therefore is a number, by definition. The underscore indicates
the sequence is of finite length.

If a number does not have a name, one can call it ($N_{1}$ union
$N_{2}$), where $N_{k}$ is named. Consider for example, the the
number 39, which an be called Thirty union Nine. Notice, that here
one is not dealing with positional value of a digit of a representation
system. 
\begin{defn}
$Natural\,\, count$ is defined in the context of constant einheit.
Consider the minimal set, which is the null set. If the elements are
ordered, call the count zero. Increment this set, by repeating the
union operation, to $\mathfrak{s=o\cup k\cup k\cup k\cup...\cup k}$.
The set of the all $\mathfrak{k}$-s has the size of $N$, (in specific
cases, same as the size of the resultant set $\mathfrak{s}$). If
elements of this set is labeled and ordered, such that the count is
$N$, then it is called a natural count. The numbers constructed,
$\forall N$, thereby are natural numbers.\end{defn}
\begin{rem}
In this definition, the size of a set is identified with the Natural
count of the einheit -s that construct the set. Natural counts are
always defined w.r.t. an einheit. A natural count $n$ with einheit
$u$ implies the possible existence of a set, which is constructed
from $\underline{\bigcup}u$, and the set of all $u$ -s used for
this construction has size $n$. Any value, while assigned an einheit,
can become a natural count. The einheit of a natural count may be
replaced with another, using certain operations.\end{rem}
\begin{defn}
An $einheit$ is a constant set, used in repeated union to construct
other sets. A $natural\,\, einheit$ is the set which has the size
one. Here the ``one'' identifies to the standard, everyday meaning
of the word ``one'', as in one llama, or one submarine or one Tom
Sawyer. 
\end{defn}

\subsubsection{Variable Einheit }

Of course the choice of einheit is arbitrary. This leads to the following
remarks:
\begin{rem}
The same number can be generated by different combinations of different
einheit. \end{rem}
\begin{proof}
Since it is a claim of a possibility, the proof can be simply illustrated
by an example. Consider the number $N=\mathfrak{k\cup k\cup...\cup k}$.
Let $\mathfrak{k=k_{x}\cup k_{x}}$. Hence $N=\mathfrak{(k_{x}\cup k_{x})\cup(k_{x}\cup k_{x})\cup...\cup(k_{x}\cup k_{x})}$.
This is a different combination, that represents the same $N$. Notice
that there is no boundary on the sequence length that constructs $N$.
QED.\end{proof}
\begin{cor}
All numbers can be represented via different combinations of different
einheits.\end{cor}
\begin{rem}
One number can be a combination of unlimitedly many different einheits.\end{rem}
\begin{proof}
From the axioms, union between any sets is possible. Hence, any sets
of any einheits exist. QED.\end{proof}
\begin{rem}
There total set of numbers is unlimitedly large.\end{rem}
\begin{proof}
$\forall\underline{\bigcup}\mathfrak{K=o_{k}},\exists\mathfrak{o_{k}\cup k|k\neq\phi}$.
Hence, it is possible to create unlimitedly many numbers. QED.\end{proof}
\begin{rem}
There exists unlimitedly many numbers between any two given numbers.\end{rem}
\begin{proof}
Assume the numbers are $\mathfrak{0}$ and $\mathfrak{0\cup K}$.
$\forall\mathfrak{K},\exists\mathfrak{K^{\prime}\subset K}.$ 

Therefore, $\forall\mathfrak{o},$ $\mathfrak{o\cup K>o\cup K^{\prime}>o}$,
and similarly, there exists a number between $\mathfrak{o\cup K^{\prime}}$
and $\mathfrak{o}$, and also between $\mathfrak{o\cup K}$ and $\mathfrak{o\cup K^{\prime}}$,
and ad infinitum. QED.\end{proof}
\begin{rem}
The set of all numbers is closed.\end{rem}
\begin{proof}
Any number, and any set, in this system is dealt as a set. That is,
what in other system is a number, in this system is a set. Singleton
elements automatically assume the status of a set. An operation on
a set produces, therefore another set. Since any set can be represented
in this system as a combination of different einheits, any set is
a valid member of the class this system represents. Therefore, the
set of all numbers (or the class of all sets) is closed, under any
operation. QED.
\end{proof}

\subsubsection{Number Line, Number Plane and Others }

The number line is an ordered sequence of all numbers generated by
all einheits, that are comparable to each other. Obviously, since
each einheit can generate infinite numbers, one line can contain only
(strictly) two types of einheits, because, it can be expanded to infinity
in two directions.

The next pair of comparable einheit can be placed along a different
line, independent to the all previous lines. Hence pairwise, the einheits
define a set of basis, which can be used to describe a $N$-dimensional
space. 
\begin{example}
The Positive and Negative Reals are comparable, under certain conditions,
and form one number line. Positive and Negative Imaginary numbers
are also under similar conditions comparable to each other, but not
comparable to Reals. Therefore, the real line and the imaginary line
define a number plane. This is the Argand Plane.

The comparison between any two points in a general $N$-dimensional
space is therefore not simple. There exist $N$ pairs of coordinates.
An order relation may, however, be constructed from a particular pair.
Therefore, the order relation is not unique.

An unknown number, is therefore a point in the space of unknown size.
The size of real numbers are therefore real size, and the imaginaries
have imaginary sizes. 
\end{example}

\subsection{Operations}

The set of ECN is a formal field, with operations defined on it. We
primarily define two order preserving operations, namely addition
and multiplication.

\subsubsection{Order Similar}

Consider an order sequence $a>b>c>...>d$, and an operation $f$,
such that $f(a,w)=p,\, f(b,w)=q,\, f(c,w)=r,\, f(d,w)=s$, for some
$w$. If $p<q<r<...<s$ or $p>q>r>...s$, then call the function an
order similar one. Given the sequence $a>b>c>...>d$, if $p>q>r...$
etc, then it is a monotonic operation, otherwise it is an antitonic
operation. 
\begin{rem}
On a given set, a monotonic operation and an antitonic operation is
isomorphic. An order similar operation implies its inverse is also
order similar. The proof is trivial.
\end{rem}
We investigate the invariant(s) of order similar operations. It is
trivially clear that
\begin{rem}
The closure, and the initial (resp. final) segments are not necessarily
invariants to a order similar operation. They are invariants for a
monotonic operation. In order theory, the monotonic operation is known
as order preserving.
\end{rem}
However, isomorphism implies, that the range and image of the operation
has the same natural count of elements. Formally:
\begin{rem}
The height, defined as the natural count of the sequence (resp. subsequence)
$-$ einheit, of the sequence (resp. subsequence) is an invariant,
under order similar operations. 
\end{rem}
Other invariants may be found from the definition of order similar
operations. Here we consider the concept of neighborhood, and introduce
the concept of zwischenraum.

Consider a sequence $S=a,b,c,...,d$, such that, $a<b<c<...<d$ apply.
Assign order labels $\alpha,\beta,\gamma,...,\delta$ on them respectively
to represent them using a graph. A subsequence $s$ of $S$ is $a,b,c$.
If we represent $s$ with a graph $g$, then it is trivially clear
that the neighbors of $b$ are $a$ and $c$. Similar statements apply
for every subsequence of length $3$ in $S$. Under an order similar
operation, $f$, $f(b)$ has the neighbors (in a graph-representation
with graph $g^{\prime}$) $f(a)$ and $f(c)$. From remark 3.13, it
is trivially clear that $g$ and $g^{\prime}$are isomorphic, specifically,
$b$ and $f(b)$ are. Hence they have isomorphic neighborhood. Same
applies for any subsequence in $S$. Hence the neighborhood matrix
is invariant. 
\begin{defn}
This permits to define a property of such $3$-tuples, which we will
call $Zwischenraum$ (from German ``space in-between'') invariance.
Consider a sequence $S:a,b,c,d,...,z$. Let $a<b<c$, such that, $\nexists d|a<d<b\vee b<d<c$.
Then , $b$ is said to occupy the $Zwischenraum$ of $a$ and $c$,
symbolically, $b\in\mathfrak{Z(}a,c\mathfrak{)}$. Clearly, $f(b)$
is occupying the zwischenraum of $f(a)$ and $f(c)$, if $f$ is an
order preserving operation. Therefore, the zwischenraum for each $2$-tuple,
constructed from an ordered sequence, if uniquely defined by a single
element $e$, the the zwischenraum of the image of the $2$-tuple
is uniquely defined with the image of $e$. This property is called
$Zwischenraum\,\, Invariance$.
\end{defn}
We try to generalize this for cases, where the zwischenraum of two
elements will consist of an ordered set $s$ consisting of more than
one elements. The zwischenraum of the image of the two bounding elements
consist of the image of $s$. This is the sought generalization. Moreover,
the set $s$ itself is also subject to zwischenraum invariance, and
so is any subset of $s$. Formally:
\begin{thm}
Any sequence with natural count larger than $2$, if admits a zwischenraum
invariance, then any subsequence of it of length having natural count
$2$ admits zwischenraum invariance to the same operation.
\end{thm}

\subsubsection{Addition and Multiplication}

\paragraph{Addition}

Addition is a monotonic binary operation, symbolized with $+$, with
following properties. 
\begin{itemize}
\item If $A<B<C<...<D$, then $A+k<B+k<C+k<...<D+k$, for all $A,B,C,...,D,k$
\item $(p+k)\backslash p$ does not depend on $p$, but on $k$, for all
$p,k$
\item $a+p$ is either $<a$ or $>a$, both may hold. This is applicable
for all $a,p$
\item Axiom of associativity and axiom of commutativity holds.\end{itemize}
\begin{lem}
The addition operation may be represented by set union.
\end{lem}
Immediately, one realizes, that there exists at least one $p,$ for
which $a+p=a$. Such $p$-s are called neutral to addition. The addition
operation, from Euclidean axioms, is comparable to set unions. Let
$\mathfrak{u}$ be the einheit. $a+p$ is defined as the size of $a\mathfrak{\cup\underline{\bigcup}u}$,
where the count of $\mathfrak{\underline{\bigcup}u}$ is $p$.

\paragraph{Multiplication}

Multiplication, an order similar operation, is similarly defined,
symbolized with $\times$ or $*$, where
\begin{itemize}
\item If $A<B<C<...<D$, then either $A*k<B*k<C*k<...<D*k$, or $A*k>B*k>C*k>...>D*k$
for all $A,B,C,...,D,k$
\item $(p*k)\backslash p$ depends on both $p$, and $k$, for all $p,k$
\item $(a*p$) is either $<a$ or $>a$, both may hold. This is applicable
for all $a,p$
\item Axiom of associativity and axiom of commutativity and axiom of distributivity
holds.
\end{itemize}
There also exists neutral to multiplication, such that $a*p=a$. Further,
$a*p$ is the size of $\mathfrak{\underline{\bigcup}a}$, where, count
of $\mathfrak{\underline{\bigcup}a}$ is $p$.

Both operations will be extended in forthcoming discussion to account
for different types of einheit.
\begin{claim}
The total set of all numbers, written $\mathfrak{n}$, is a group.\end{claim}
\begin{proof}
$\mathfrak{n}$ is a set, with two operations defined on it, and is
closed, by definition, and from remark 3.11 to both. Both operations
agree to the axiom of associativity, and assume an identity element.
In next sections, the inverses will be discussed. Hence group axioms
are fulfilled. QED.
\end{proof}

\subsubsection{Operands of $+$ and $\times$}

Above the operations are described in terms of Order Labels, which
are sizes of sets. The size of sets are identified with the natural
count of einheits that construct the set. 

Sets need not be collection of disjoint elements, the elements may
touch (in a topological sense) or overlap each other. In practicality,
it is more often the case than not, that, the elements are not disjoint. 

The addition operation deals with the size of the sets, which is intuitively
clear, and is supported by Euclidean axioms. If both operands possess
the same einheit, then the result also has the same einheit as well.
Otherwise, either one einheit is transformed (we will use the term
transcribed, i.e. represented differently, instead) to the other,
and then the result is computed, or they are left as disjoint union. 
\begin{example}
Assume the addition between $a$ and $b$ is performed (which is the
union of two underlying sets having natural counts $a$ and $b$),
where $a$ is imaginary, and the other is real (the imaginaries are
not yet defined explicitly). Recall each natural count is defines
w.r.t. a constant einheit. In this particular case the sum is left
as $a+b$, because the real and the imaginary einheit is separate.
Had they both the same einheit, then the result would be represented
by $c$, which is the natural count of the set $A\cup B$, $A$ and
$B$ having natural counts $a$ and $b$ respectively. The natural
count $c$ will be defined w.r.t. the same einheit as both $a$ and
$b$.
\end{example}
The multiplication operation is however, slightly more complex. If
the operation is performed on two operands, with two (not necessarily
same) einheits, for example to calculate $a\times b$, a set $A$
set with size $a$ is constructed, w.r.t the einheit of $a$. $b$
is now defined as a natural count, w.r.t. its own einheit. A sequence
$S$, each of its element being $a$, is constructed with count $b$.
The count labeling need not be dependent on the size. Another set
$R$ is constructed by repeated union of elements of $S$. Then $R$
is the result of the multiplication operation.

\subsubsection{Further Operations on $\mathfrak{n}$ }

Although the purpose of $\mathfrak{n}$ here is to serve simply as
a tagging system, that permits somewhat flexible re-tagging system,
and possibility to generate new numbers between any two in a comprehensive
way. It is now of interest, to investigate, if $\mathfrak{n}$ assumes
a algebraic structure that can be exploited.

\paragraph{Length of a Line Segment}

Given two points $\mathfrak{p_{1}}$ and $\mathfrak{p_{2}}$ in the
number line $\mathfrak{l}$, where $\mathfrak{p_{1}<p_{2}}$ holds,
assume there exists a third number $\mathfrak{p_{3}|p_{1}+p_{3}=p_{2}}$,
where the $+$ operation is defined above. Since $\mathfrak{p_{3}}$
is also a number, it exists in $\mathfrak{n}$. $\mathfrak{p_{3}}$is
called the length of line segment between $\mathfrak{p_{1}}$ and
$\mathfrak{p_{2}}$.

Now, to calculate $\mathfrak{p_{3}}$, we need $\mathfrak{\backsim p_{1}}$,
the inverse of $\mathfrak{p_{1}}$w.r.t addition with $\mathfrak{p_{2}}$.
We have $\mathfrak{p_{1}}+\mathfrak{p_{3}}=\mathfrak{p_{2}}$. From
previous discussion, the operands are sizes, hence the Euclidean axioms
apply. Hence we add $\mathfrak{\backsim p_{1}}$ to both sides. RHS
is the $\mathfrak{p_{3}+}$ neutral to addition, which is just $\mathfrak{p_{3}}$
and LHS is $\mathfrak{p_{2}+(\backsim p_{1})}$. 

This discussion leads to the length operation. The length operation
is written as $|a-b|$. It is defined as $a+(\backsim b),\,\, iff\,\, a<b,b+(\backsim a)\,\, otherwise$. 
\begin{rem}
The outcome of this operation has the same einheit as the operands.
\end{rem}
The definition of a subtraction operation follows immediately. The
Euclidean axioms uses the notion of ``taking away'', which is comparable
to set exclusion operation. Translated to notion of order labels,
$a-b$ , where the $-$ symbol designates the subtraction. 
\begin{rem}
From Euclidean axioms, subtraction is only then defined, when $a\not>b$. 
\end{rem}
$a-b$ therefore is the size of the set $A\backslash B$, where $A$
has the size $a$, and $B$ the size $b$. If the under-laying sets
have similar einheit, then the result is also the same unity. 

For any $a$, it is immediately clear, that $a-a=0$, the size of
the null set. The $0$ is again the neutral to addition. So, $a-a=0=a+\backsim a$.
Now, again, from Euclidean axioms, adding any $b$ to both sides,
preserve the equality relation. Therefore, the subtraction operation
is same as adding the inverse.

\paragraph{Negative Numbers and Extension}

A negative numbers may be treated as the inverse to a given number
in a group under addition. The definition is rather synthetic, that
it does not immediately follow from Euclidean axioms. Inverses of
negatives ($(a^{-1})^{-1}=a$ holds) are called positives. 
\begin{claim}
Each set has an inverse of it, such that set $+$ inverse of it $=\phi$.
Those inverses follow the Euclidean axiom. These inverses may be just
mathematical concepts.\end{claim}
\begin{prop}
In this context, we define the the order relation as follows. Let
$\backsim a$ and $\backsim b$ be two negative numbers. $\backsim a>\backsim b$
$iff$ $a>b$. \end{prop}
\begin{defn}
Call the numbers with natural einheits $Universal\,\, Positives$,
and the corresponding negatives $Universal\,\, Negatives$. \end{defn}
\begin{rem}
In particular einheit scheme, negatives has the same einheit as the
corresponding positives, then they are inverses to each other, and
results in a neutral, under addition.\end{rem}
\begin{prop}
Sometimes, it is necessary to order the positives and the negatives
(or other pairs) in the same sequence. For that purpose, the number
line is constructed. 
\end{prop}
Universally, assign $\sim a<a$. Start $\sim0$, and continue till
$\sim\infty$, then the sequence is $\sim0<0<...<\sim a<a<...<\sim\infty<\infty$.
This ordering permits a semi infinite line to be constructed. If,
however, $\backsim a>\backsim b$ $iff$ $a<b$ is used, together
with $\sim a<0$ if $\sim a$ is a universal negative, then we arrive
at the usual positive-negative ordering.

\paragraph{Division as Multiplication with Inverses }

Division operation is defined for multiplication. If $a\times b=$
neutral of multiplication, then, $c\div a=c\times b$. Division operation
is therefore also order similar.

The division operation splits the set $c$ in parts, the set of which
admits a count. The count is $a$.

\subsubsection{The Algebraic Structure of $\mathfrak{n}$}

\paragraph{Quasigroups of $\mathfrak{n}$}
\begin{thm}
A subset $\mathfrak{s}\subset\mathfrak{n}$ that contains all numbers
of same einheit, and only numbers of same einheit is a group, under
the addition operation.\end{thm}
\begin{proof}
Assume $a,b\in\mathfrak{s}$. The addition operation results in $c$.
$c$ has the same einheit as $a$ and $b$. Therefore, $c\in\mathfrak{s}$.
this holds $\forall a,b\in\mathfrak{s}$. Therefore $a,b\in\mathfrak{s}\implies a+b\in\mathfrak{s}$.
Hence $\mathfrak{s}$ is a magma.

$\forall a,c$, we have $b=c-a$. Note that although the subtraction
operation requires the presence of an inverse, we do not require the
inverse to be present in $\mathfrak{s}$ itself. $\mathfrak{s}$ is
therefore a quasigroup.

$0$ is the element which is the neutral to addition, and $0\in\mathfrak{s}$.Therefore
$\mathfrak{s}$ is a loop.

Associativity of addition holds. Hence $\mathfrak{s}$ is a group.
\end{proof}
This theorem can be trivially extended to :
\begin{thm}
A subset $\mathfrak{s\in n}$ which contains all numbers of a particular
einheit, and all it's inverses, and only the mentioned numbers, is
a group.
\end{thm}
Then we consider the segment length operation. The length segment
operation in general is not associative, nor does it admit an unique
definition of division. Therefore it does not admit a group-like structure.

\paragraph{Construction of a Finite Group }

Given $a,b,c\in\mathfrak{s}\subset\mathfrak{n}$, we consider if $\mathfrak{s}$
can be constructed in a group, based on the order properties. The
addition operation automatically admits a group structure. 

The multiplication operation depends on the assignment of count. It
can be trivially shown that :
\begin{thm}
If the subset $\mathfrak{s}\subset\mathfrak{n}$ contains only positives,
then it admits a group structure. If the subset contains only negatives,
then assigning negative counts to multiplication admits a group structure.
If $\mathfrak{s}\subset\mathfrak{n}$ contains both positives and
negatives, then under the usual multiplication scheme, it admits a
group structure. Same applies for division operation.
\end{thm}
Now consider the transformation group containing the operations defined
above. Trivial investigation shows:
\begin{thm}
The sets of operations addition, multiplication (division), respectively,
and count assignment all admit a group structure.
\end{thm}
Obviously each of them are infinite groups. The composition is simply
the convolution of two operations. Such groups are extremely simple
in structure, and of less interest in this context.

We next consider groups whose elements take the form $\lightning\mathfrak{a}$,
where $\lightning$ is a operation, and $a$ is a number. If $\lightning$
is the addition operation, then the structures are similar to what
is seen with the addition group already.

However, interesting observations can be made if $\lightning\colon\times$.
This groups contains a subgroup, which is constructed of non-prime
multiples of natural einheits and their inverses. This subgroup is
also a normal subgroup. A series of quotient groups to this is immediately
generated. Each of them is also abelian. 

These groups however does not admit a soluable structure. Nonetheless,
we consider the order properties.
\begin{thm}
Each such normal subgroup $\mathfrak{p}\triangleleft\mathfrak{n}$
is $\mathfrak{n}$ orderable, then they admit an order which is invariant
under $n\mathfrak{p}n^{-1}\forall n\in\mathfrak{n}$
\end{thm}
The proof is trivial. Further investigation shows $\mathfrak{p}$
does not contain any subgroup.

\subsection{Generation of Numbers and Different Einheits}

Armed with the definitions of numbers and usual operations, the next
section is dedicated to the original question.

\subsubsection{Generation}

Assuming two non-neighboring elements are tagged, of extreme interest
is to find a method to tag the elements in between the tagged ones.
For that, the set inbetween the two tagged elements is split, at will,
and split sequence is assigned with a count.

Therefore, there are two steps inherent to the process. The splitting
operation and the count assignment operation, from theorem 3.5 assumes
a structure of a finite group.

Formally, assume a chain $\mathfrak{a_{1}\in a_{2}\in a_{3}\in...\in a_{n}}$.
Assume further that $\mathfrak{b_{1},b_{2}\in a_{n}}$ are tagged,and
they are NOT neighboring elements. Then, we want to tag all elements
in $\mathfrak{a_{1}}$. 

Trivially, and using the discussions above, it is clear that sets
may be tagged independent to each other.

\paragraph{Uniqueness Requirements}

We consider the uniqueness of each tagging. Clearly, each $\mathfrak{a_{i}}$
has a different einheit. However, the tagging of objects between $\mathfrak{b_{a},b_{b}\in a_{i}}$
is definitely different than $\mathfrak{b_{a},b_{b}\in a_{j}}$. Therefore
$\mathfrak{b_{a},b_{b}\in a_{i}}$ must be different numbers than
$\mbox{\ensuremath{\mathfrak{b_{a},b_{b}\in a_{j}}}}$. This difference
is created by different einheits of $\mathfrak{a_{i}}$and $\mathfrak{a_{j}}$. 

Assume $\mathfrak{b,c\in a_{i}}$ are tagged. $\mathfrak{a_{i}}$
has einheit $\mathfrak{u_{i}}$. Splitting $\mathfrak{b-c}$ results
in einheit $\mathfrak{u_{j}}$. Then $\mathfrak{d|b<d<c}$ has einheit
$\mathfrak{u_{i}+u_{j}}$.

\subsubsection{Construction of Finite Group and Consequences}

We consider the division (splitting) operations that take the form
$\times\frac{1}{p}$, where $p$ is a number with a natural, positive,
einheit.

It has been shown in the previous discussion that if $p$ is not a
prime number, then the set of the said operations form a normal subgroup,
and there exists quotient groups to this. 

Therefore, the tagging algorithm can take the advantage of creating
a quotient group, which still preserves the ordering, the einheit,
yet accommodates for newer symbols than the original group. That is,
the measure of $\mathfrak{b-c}$ is changed, and is a natural multiple
of of the original value. If the original tagging had a natural count,
is not however preserved. Therefore, a new natural count can be assigned
with elements, making place inserting new symbols. One can trivially
show that:
\begin{thm}
(An operation from) The quotient group preserves the order. If a triple
is considered, with the initial elements that serve as the boundary,
and the set whose elements are to be ordered in between, then this
triple is also preserved.
\end{thm}
Since there exists normal subgroups, and the elements $\mathfrak{a_{i}}$-s
of the subset chain can be accessed independently to each other, the
splitting operation, as the tagging also, can be done independent
to each other. The result can be uniquely determined if the respective
einheit is known.

Ordering of numbers having composite einheit in form $\mathfrak{u_{1}+u_{2}+...+u_{n}}$
is done as follows. Rewrite the two comparable numbers as $\mathfrak{s_{1}u_{1}+s_{2}u_{2}+...s_{n}u_{n}}$,
such that $\mathfrak{v:u_{1}>u_{2}>...>u_{n}}$ holds. Then let $\mathfrak{u_{i}}$
be the first einheit in $\mathfrak{v}$, where the corresponding $\mathfrak{s_{i}}$
is different in the two numbers. Order the $\mathfrak{s_{1}}$, that
is the order of the two numbers

\section{Solution to the Original Problem and Application}

\subsection{Solution of the McDougal Problem }

Now, the McDougal cave problem admits a trivial solution. Attach a
different einheit to each cave and chamber and tunnel. Require the
einheit corresponding to one object that is included in another be
lesser than that of later. If an object, inbetween two is to be tagged,
while the both of them are already tagged, then tag them with a number
generated from that of the former two.

The application of this system finds use in controlling the execution
of any computer simulation, and in database manipulation. Indeed it
is possible to tag a sequence of similar objects in McDougal cave
as 21,22,23,...35... etc, but the change of the second digit from
right may make the fact less obvious that the objects are actually
the same. Therefore proposed tagging is 2$\oplus$1,2$\oplus$2,...,2$\oplus$267,...
etc, where $\oplus$ designates the addition of different einheits,
and the tagging implies that all of them belongs to the zwischenraum
between 2 and 3, (with einheit larger than the objects in consideration).
If there are $n$ such object, then the tag for the $m$-th one can
be generated from $2,3,\{n,m\}$ using the method illustrated above.

If, a new object between two tagged one is to be inserted, then, the
problem is a bit complex, because $m$ is a natural count. Therefore,
a re-tagging is necessary. However, a quotient group sorts the issue
- for example, you can insert an object in the second place, in the
sequence, 2$\oplus$1, 2$\oplus$2,3$\oplus$1,3$\oplus$3, by re-tagging
the first and second, to 2$\oplus$1 and 2$\oplus$3, where the second
tagging is an operation in a quotient group of the group in which
the first tagging belongs. The quotient group preserved the ordering,
and the terminal tags. This greatly simplifies the process for block
based heuristic simulation algorithms.

\subsection{Sorting}

We consider the sorting problem. Here, sorting is done heuristically,
and an element from the unsorted sequence is inserted in the sorted
sequence, in it's place.

The algorithm measures the rate of change of values as it progress
along the sequence starting from a particular boundary. Then it predicts
a location where a new, known element can be inserted. If the insertion
does not disrupt the order, then insertion is followed by algorithm
termination. Otherwise, taking the predicted location as a boundary,
the algorithm repeats itself. 

Infinity loops are not avoidable. In case of an infinity loop, the
algorithm fails.

This algorithm, is works at the worst case, is of complexity $O(n)$.

\subsection*{References:}

{[}1{]}: Gonsor, H. An introduction to the theory of Surreal Numbers,
Cambridge University Press, 1986
\end{document}